\documentclass[preprint,12pt]{elsarticle}
\usepackage{graphicx}
\usepackage{amssymb}
\usepackage{amsmath}

\begin{document}

\begin{frontmatter}

\title{On the width of the $K^-$D atom ground state} 

\author[a]{N.~Barnea} 
\author[a]{E.~Friedman} 
\author[a]{A.~Gal\corref{cor1}} 
\cortext[cor1]{Corresponding author: A.~Gal, avragal@savion.huji.ac.il} 
\address[a]{Racah Institute of Physics, The Hebrew University, 91904
Jerusalem, Israel} 

\begin{abstract} 
Experiments at DA$\Phi$NE-Frascati and at J-PARC are scheduled to produce 
$K^-$D atoms and observe their X-ray cascade down to the 1$S$ ground state 
(g.s.), thereby measuring its strong-interaction width and shift away from 
a purely Coulomb state. A width $\Gamma_{1S}\lesssim 1$~keV will ensure good 
resolution of the X-ray transitions feeding the 1$S$ g.s. Here we study 
the expected $K^-$D 1$S$ g.s. width from the perspective of global fits 
to level shifts and widths in heavier kaonic atoms across the periodic table, 
using $K^-$ nuclear optical potentials constructed from $\bar K N$ chiral 
interaction models. Special attention is paid to the subthreshold energy 
at which the $\bar K N$ subsystem interacts in the $K^-$D atom g.s. Within 
this approach we predict strong-interaction upward level shift of close to 
700~eV and width of about 1.2 to 1.3~keV for the $K^-$D atom 1$S$ g.s., in 
fair agreement with genuinely three-body $K^-$D atom calculations. Comparison 
is made with $\pi^-$D atom phenomenology. 
\end{abstract} 


\end{frontmatter}

\section{Introduction} 
\label{sec:intro} 

Strong-interaction level shifts and widths observed in hydrogen and deuterium 
hadronic atoms provide valuable information on hadron-nucleon scattering 
lengths, as accomplished for pions and antiprotons~\cite{Got04}. For $\bar K$ 
mesons, following several high-resolution experiments~\cite{Cur19}, the 
SIDDHARTA-1 experiment at DA$\Phi$NE-Frascati~\cite{Baz11,Baz12} provides 
a fairly accurate determination of the $K^-p$ complex scattering length by 
measuring the strong-interaction level shift and width in the $K^-$H atom 
1$S$ g.s. To constrain the $K^-n$ complex scattering length one needs to 
form $K^-$D atoms and observe their X-ray cascade down to the 1$S$ g.s. Two 
experiments, SIDDHARTA-2 at DA$\Phi$NE-Frascati and E57 at J-PARC are running 
or scheduled to run, aiming to observe this X-ray cascade~\cite{Cur19}. It is 
tacitly assumed, based on past theoretical calculations, that the broadening 
of $K^-$D atom cascade lines leading to the 1$S$ g.s. is of order 1~keV. 
Indeed the recent genuinely three-body $K^-$D atom calculations listed in 
Table~\ref{tab:calcs} support this working assumption. In these calculations 
the $K^-p$ input interaction reproduces the SIDDHARTA-1 $K^-$H 1$S$ 
level shift and width within their measurement uncertainties, whereas 
the weaker $I=1$ $K^-n$ input interaction remains largely model dependent. 
We note that the $K^-$D 1$S$ level shift and width result in these 
works {\it directly} from the calculated $K^-$D atom complex energy g.s. 
eigenvalue without using the Deser formula~\cite{DGBT54} or its improved 
versions~\cite{T61,CRW92,MRR04,BER09}.{\footnote{The Deser formula links 
the $K^-$D atom 1$S$ g.s. level shift and width to the complex scattering 
length $a_{K^-d}$, and is known to invoke errors of up to 30\% for the width 
$\Gamma$~\cite{Shev14}. Following Ref.~\cite{Got04} we use D for deuterium 
atoms and $d$ for deuteron nuclear attributes.}}  

\begin{table}[htb]
\caption{$K^-$D 1$S$ level shift ($\epsilon_{1S}$) and width ($\Gamma_{1S}$) 
in eV from three-body calculations. Note that $\epsilon_{1S} < 0$ stands for 
upward shift with respect to the purely Coulomb case.} 
\begin{center} 
\begin{tabular}{cccc} 
\hline 
$-\epsilon_{1S}$ & $\Gamma_{1S}$ & Method & Reference \\ 
\hline 
828&1055& Faddeev + $V_{\rm opt}$ & Shevchenko-R\'{e}vai 2014~\cite{Shev14} \\ 
800$\pm$30 & 960$\pm$40 & Faddeev & R\'{e}vai 2016~\cite{Rev16}; see also 
\cite{DRS15}  \\ 
670 & 1016 & 3-body Schroedinger & Hoshino et al. 2017~\cite{Hosh17}  \\ 
\hline 
\end{tabular} 
\end{center} 
\label{tab:calcs} 
\end{table} 

Given the model dependence of the input subthreshold $\bar K N$ interaction, 
in the presence of a quasi-free spectator nucleon in Faddeev three-body 
calculations, the agreement to $\lesssim$10\% between all values listed in 
Table~\ref{tab:calcs} for the $K^-$D 1$S$ g.s. width, $\Gamma_{1S}(K^-{\rm D})
\approx 1$~keV, is striking. The sensitivity of the calculated width to 
varying the poorly known real part of the $K^-n$ input interaction was 
explored in Ref.~\cite{Hosh17}, concluding that $\Gamma_{1S}(K^-$D) remains 
stable around 1 keV within less than 40~eV decrease from the value listed in 
the last line of Table~\ref{tab:calcs}. The sensitivity to the imaginary part 
of the $K^-n$ input interaction can be inferred qualitatively by comparing the 
values of $\Gamma_{1S}(K^-$D) from Refs.~\cite{Shev14,Hosh17} as listed in the 
table, resulting in a $\sim$40~eV increase of the calculated $\Gamma_{1S}
(K^-$D) with respect to Ref.~\cite{Hosh17}. 

These $K^-$D calculations and many other published ones were implicitly 
questioned recently by Liu et al.~\cite{LWLT20} in the context of 
multiple-scattering expansions of the $K^-d$ scattering length in terms of 
$K^-p$ and $K^-n$ subthreshold scattering amplitudes. Shifting the $\bar K N$ 
energy argument from threshold towards the $\Lambda$(1405) resonance, to 
accommodate the recoil energy of the spectator nucleon, is claimed there to 
increase $\Gamma_{1S}(K^-$D) from $\approx$1.1~keV up to $\approx$2.3~keV. 
However, recalling that a recoil energy shift is implemented standardly in the 
Faddeev calculations listed in Table~\ref{tab:calcs}, the meaning of such 
a claim is questionable. Here we consider this issue differently, guided 
by lessons from optical model studies of kaonic atoms strong-interaction 
level shifts and widths across the periodic table~\cite{FG12,FG13,FG17}. 
The optical model $K^-$D calculations with subthreshold $K^-N$ input 
reported in the present work do not support Liu et al.~\cite{LWLT20}, 
yielding $\Gamma_{1S}(K^-$D)$\approx$1.2$-$1.3~keV, in fair agreement 
with the values listed in Table~\ref{tab:calcs}. 

The paper is organized as follows. In Sect.~\ref{sec:sub} we review the issue 
of subthreshold energy used in $K^-$ atoms calculations, particularly in 
few-body systems, applying it in Sect.~\ref{sec:KbarD} to optical model 
calculations of the $K^-$D atom 1$S$ g.s. shift and width and the $K^-d$ 
scattering length. A similar calculation of the $\pi^-$D atom for which the 
1$S$ g.s. level is known experimentally is reported in Sect.~\ref{sec:piD}, 
thereby testing the optical model methodology of the present work. In 
Sect.~\ref{sec:MS} we comment on multiple-scattering expansions of the $K^-d$ 
scattering length in terms of $K^- p$ and $K^- n$ scattering amplitudes input, 
demonstrating the extent to which the results of Liu et al.~\cite{LWLT20} 
differ from those of a genuinely three-body Faddeev calculation~\cite{TGE81}. 
Finally, conclusions drawn from the present work are summarized in 
Sect.~\ref{sec:concl}.

\section{Subthreshold energy considerations in $K^-$ atoms} 
\label{sec:sub} 

The $K^-$-nucleus optical potential used in global fits to strong-interaction 
data in kaonic atoms from $^7$Li to $^{238}$U is of the form $V_{K^-}=
V^{(1)}_{K^-}+V^{(2)}_{K^-}$~\cite{FG17}. Its single-nucleon term 
$V^{(1)}_{K^-}$ is given by 
\begin{equation} 
2{\mu_K}V^{(1)}_{K^-}(\rho) = -4\pi\,({\tilde f}_{K^-p}(\rho)\rho_p + 
{\tilde f}_{K^-n}(\rho)\rho_n), 
\label{eq:V1} 
\end{equation} 
where $\mu _K$ is the $K^-$-nucleus reduced mass. The in-medium $K^- N$ 
scattering amplitudes in the $K^-$-nuclear center-of-mass (cm) frame, 
${\tilde{f}}_{K^-N}(\rho)$, are kinematically related to the in-medium 
$K^- N$ cm amplitudes $f_{K^-N}(\rho)$, 
\begin{equation} 
{\tilde{f}}_{K^-N}(\rho)=(1+\frac{A-1}{A}\frac{\mu_K}{m_N})\,{f}_{K^-N}(\rho), 
\,\,\, N=p,n.  
\label{eq:f1} 
\end{equation} 
In Eqs.~(\ref{eq:V1},\ref{eq:f1}), the density dependence of $f_{K^-N}$ arises 
partly by introducing a density-dependent subthreshold energy shift $\delta
\sqrt{s(\rho)}$, defined below, and partly from Pauli correlations. The 
latter do not enter $s$-shell nuclear optical potentials and are suppressed 
here in deuterium. 

The $V^{(2)}_{K^-}$ term of the optical potential is a phenomenological 
density-dependent term representing $K^-$ multinucleon processes: 
\begin{equation} 
2\mu_K V^{(2)}_{K^-}(\rho)=-4\pi B\,(\frac{\rho}{\rho_0})^{\alpha} \,\rho, 
\,\,\,\,\,\,\,\,\, \rho_0=0.17\,{\rm fm}^{-3},
\label{eq:V2} 
\end{equation}
with a complex strength $B$ and a positive exponent $\alpha$, the three of 
which serve as fit parameters. For $\alpha=1$, as used here, the $\rho ^2$ 
dependence of $V^{(2)}_{K^-}(\rho)$ agrees with the traditional $\rho^2$ 
dependence motivated in mesic atoms by absorption on two nucleons~\cite{FG07}. 
This term accounts for more than 20\% of the $K^-$ absorption width measured 
in kaonic atoms, and it is essential in reaching good agreement between 
calculations and experiment~\cite{FG17}. 

Given that the $K^-N$ scattering amplitude (\ref{eq:f1}) is strongly energy 
dependent, owing to the $\Lambda$(1405) resonance, one needs to determine 
$K^-N$ subthreshold energy values at which $f_{K^-N}$ should enter 
$V^{(1)}_{K^-}$. The Mandelstam variable $\sqrt{s}=\sqrt{(E_{K^-}+E_N)^2 
-({\vec p}_{K^-}+{\vec p}_N)^2}$ which reduces to $(E_{K^-}+E_N)$ in the 
$K^-N$ two-body cm system is an acceptable choice, although it is not a 
conserved quantity in the $K^-$ nuclear problem. Since spectator nucleons 
move the interacting $K^- N$ two-body subsystem outside of its cm, 
(${\vec p}_{K^-}+{\vec p}_N$) no longer vanishes, leaving ${\vec p}_{K^-}$ 
and ${\vec p}_N$ little correlated and making $f_{K^-N}$ and thereby 
$V^{(1)}_{K^-}$ momentum dependent, or equivalently density dependent as 
discussed and practised in past $K^-$ atoms studies~\cite{FG12,FG13,FG17}. 

In $K^-$ nuclear few-body problems, $A\leq 4$, where density is 
not introduced explicitly, one requires instead that the chosen 
$\delta\sqrt{s}=\sqrt{s}-\sqrt{s_{\rm th}}$ input value is reproduced by its 
expectation value $\langle\,\delta \sqrt{s}\,\rangle$ output, expressed as 
\begin{equation} 
\langle \delta\sqrt{s} \rangle = -\frac{{\cal B}_{A}}{A}-\xi_{N}\frac{1}{A}
\langle T_A \rangle +\frac{A-1}{A} {\cal E}_{K^-} -\xi_A \xi_{K^-}
\left ( \frac{A-1}{A}\right )^2\langle T_{K^-} \rangle, 
\label{eq:deltas} 
\end{equation}
where the right-hand side is generated by solving the corresponding 
hadron-nuclear few-body problem \cite{BGL12,BFG17,BBFG17}. Here, $T_A$ 
and $T_{K^-}$ denote the nuclear and $K^-$ kinetic energy operators in 
appropriate Jacobi coordinates, ${\cal B}_A$ is the total binding energy 
(including $K^-$ if bound) and ${\cal E}_{K^-}=\langle H-H_A \rangle$ with 
each Hamiltonian defined in its own cm frame. The constants $\xi$ are defined 
through $\xi_{N(K^-)}=m_{N(K^-)}/(m_N+m_{K^-}),~\xi_A=Am_N/(Am_N+m_{K^-})$.

\section{Application to $K^-$D atoms} 
\label{sec:KbarD} 

Applying Eq.~(\ref{eq:deltas}) directly to $K^-$D atoms, with $A=2$, 
${\cal B}_d$=2.2~MeV and $\langle T_d\rangle \approx 15-20$~MeV, we disregard 
the two $K^-$ terms which are negligibly small in light $K^-$ atoms. The 
resulting $K^-$D atom subthreshold downward energy shift is rather small: 
\begin{equation} 
-\langle\delta\sqrt{s}{\rangle}_{K^-{\rm D}}\approx 6.0 - 7.7~{\rm MeV}, 
\label{eq:deltasKD} 
\end{equation} 
considerably smaller than the several tens of MeV range of values encountered 
in heavier kaonic atoms analyses~\cite{FGCHM18}. Below we use a representative 
value of $\delta\sqrt{s}=-7$~MeV for demonstration. 

\begin{figure}[!h] 
\begin{center} 
\includegraphics[height=70mm,width=0.70\textwidth]{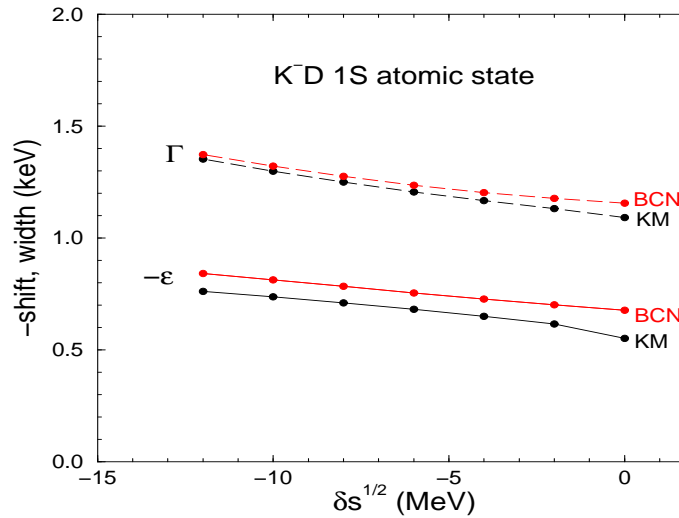} 
\caption{Calculated $K^-$D 1$S$ level shift $\epsilon_{1S}$ and width 
$\Gamma_{1S}$ as a function of the subthreshold energy $\delta\sqrt{s}$ at 
which the KM~\cite{IHW11} and BCN~\cite{BCN20} $K^-N$ amplitudes, input to 
$V^{(1)}_{K^-}$, are evaluated. The calculations use an optical potential 
$V_{K^-}=V^{(1)}_{K^-}+V^{(2)}_{K^-}$, with $V^{(2)}_{K^-}$ from global fits 
to kaonic atoms level shifts and widths across the periodic table \cite{FG17}.} 
\label{fig:subth} 
\end{center} 
\end{figure} 

The $K^-$D atom 1$S$ g.s. level shift and width are derived from the complex 
binding energy obtained by solving the Klein-Gordon (KG) equation, using an 
optical potential of the form Eqs.~(\ref{eq:V1}-\ref{eq:V2}) with Gaussian 
deuteron densities discussed in Appendix A. The $V_{K^-N}^{(1)}$ input $K^- N$ 
subthreshold amplitudes $f_{K^-N}$ are taken from the Kyoto-Munich (KM) chiral 
model~\cite{IHW11}, and from the Barcelona (BCN) chiral model~\cite{BCN19} 
(augmented recently by $K^-NN$ absorption terms evaluated in nuclear matter 
\cite{BCN20}). The $V_{K^-N}^{(2)}$ complex parameter $B$, for $\alpha =1$, 
was derived in a global fit to $K^-$ atom strong-interaction data across the 
periodic table~\cite{FG17}. Calculated 1$S$ level shift ($\epsilon_{1S}$) and 
width ($\Gamma_{1S}$) are shown in Fig.~\ref{fig:subth} as a function of the 
subthreshold energy shift $\delta\sqrt{s}$ at which the $K^- N$ amplitudes 
$f_{K^-N}$ were evaluated. The plotted values of the total width $\Gamma_{1S}$ 
are seen to increase steadily with $-\delta\sqrt{s}$, reflecting the increased 
Im$\,f_{\bar K N}^{I=0}$ as one approaches the peak of the $\Lambda$(1405) 
subthreshold resonance. However, the increase from $\Gamma_{1S}(\delta
\sqrt{s}=0)\approx 1.1-1.2$ keV to $\Gamma_{1S}(\delta\sqrt{s}=-7$~MeV)$\,
\approx 1.2-1.3$ keV is rather slow. As for absolute values we note 
that $\Gamma_{1S}(\delta\sqrt{s}=0)$=1091~eV, calculated here using the KM 
chiral model input, fares well with the value $\Gamma_{1S}$=1016~eV from 
Table~\ref{tab:calcs} derived by solving the three-body $K^-$D Schroedinger 
equation~\cite{Hosh17}, also with $K^-N$ threshold amplitudes and using 
essentially the same KM chiral model input. 

Regarding the multinucleon absorption partial width contributed by Im$\,B$, 
it comes out relatively small in these optical-model calculations, about 
80~eV in model KM and somewhat larger in model BCN. Old bubble chamber 
measurements constrain this partial width to 1.2$\pm$0.1\% of the total $K^-$D 
1$S$ width~\cite{VB70}, about 15 eV here, much smaller than our derived value 
of about 80 eV. This suggests that multinucleon absorption in heavier $K^-$ 
atoms involves $I=1$ $NN$ pairs rather than the $I=0$ $pn$ pair of which the 
deuteron consists, quite differently from the way multinucleon absorption 
in pionic atoms is perceived, as discussed briefly in the next section. 
Interestingly, the widths calculated upon setting $B=0$ exceed by about 
100~eV those for $B\neq 0$ when taken from global $K^-$ atoms fits. 
This somewhat unexpected behavior is representative of the saturation 
property of atomic widths caused by the strong imaginary part of the $K^-$ 
nuclear optical potential which suppresses the overlap of the $K^-$ atomic 
wavefunction with the nuclear density, thereby reducing its absorptive 
effect~\cite{FG99a,FG99b}. A strong Im$\,V_{K^-}$ is also responsible for 
the increased {\it repulsive} level shift $\epsilon_{1S}$ exhibited in 
Fig.~\ref{fig:subth} in spite of Re$\,V_{K^-}$ becoming more attractive 
as the $\Lambda$(1405) is approached. 

\begin{figure}[!t] 
\begin{center} 
\includegraphics[height=70mm,width=0.70\textwidth]{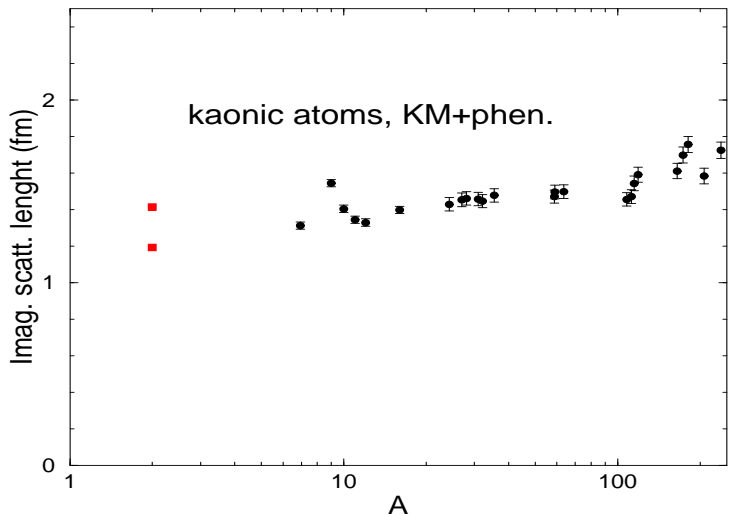} 
\caption{Im$\,a_{K^- A}$ from a global $K^-$ nuclear optical potential 
$V_{K^-}=V_{K^-}^{(1)}+V_{K^-}^{(2)}$ fit to kaonic atoms level shifts and 
widths~\cite{FG17} plotted (in black) as a function of the atomic number $A$. 
The input to $V_{K^-}^{(1)}$ are density dependent $\bar K N$ subthreshold 
amplitudes based on the KM chiral model \cite{IHW11} whereas $V_{K^-}^{(2)}$ 
is determined by the fit. The $A=2$ points (in red) are from the present 
$K^-d$ calculation (lower: $\delta\sqrt{s}=0$, upper: $\delta\sqrt{s}=-7$ 
MeV).}  
\label{fig:sat} 
\end{center} 
\end{figure} 

Further support for the saturation property of $K^-$ atomic widths is provided 
by the approximate $A$-independence found for the imaginary part of the $K^-$ 
nuclear scattering length of the $K^-$ nuclear optical potential $V_{K^-}$, 
Eqs.~(\ref{eq:V1}-\ref{eq:V2}), in a global fit to $K^-$ atoms data across the 
periodic table. This is demonstrated in Fig.~\ref{fig:sat} by the black points 
calculated using the KM chiral model $K^-N$ amplitudes~\cite{IHW11} with 
energy argument shifted in a self consistent procedure to density dependent 
subthreshold energies $\delta\sqrt{s(\rho)}$~\cite{FG12,FG13,FG17}. Applying 
the fitted $K^-$ optical potential to $K^-d$, at the $K^-N$ threshold and at 
7~MeV below it, we get the following values for the $K^-d$ scattering length:  
\begin{equation} 
a_{K^-d}:\,\,\,-0.978+{\rm i}\,1.193\,\,(\delta\sqrt{s}=0),\,\,\,\,\,\
-1.260+{\rm i}\,1.414\,\,(\delta\sqrt{s}=-7),
\label{eq:Ima} 
\end{equation} 
with $\delta\sqrt{s}$ in MeV and $a_{K^-d}$ in fm. Both imaginary values, 
added in red in Fig.~\ref{fig:sat}, compare well with Im$\,a_{K^-d}=1.32$~fm 
from the $K^-d$ Faddeev calculation using chiral interactions input reported 
in Ref.~\cite{Shev14}. Clearly they are far off the value Im$\,a_{K^-d}=2.70
$~fm claimed by Liu et al.~\cite{LWLT20}.

\section{Comparison with $\pi^-$D atoms} 
\label{sec:piD}

To make sure that applying a fitted $K^-$ nucleus optical potential to as 
light kaonic atom as $K^-$D makes sense, we followed this same methodology 
in the $\pi^-$D atom. The $\pi^-$ nuclear optical potential parameters are 
well fitted to pionic atoms data across the periodic table as demonstrated 
in Ref.~\cite{FG14}. We used this optical potential in a calculation of the 
strong-interaction $\pi^-$D 1$S$ g.s. level shift and width. No $\pi^-N$ 
subthreshold energy shift was applied since the energy dependence of the 
input $\pi^-N$ scattering amplitudes near threshold is known to be 
negligible~\cite{SAID06}. Furthermore, the $p$-wave part of the $\pi^-$ 
optical potential proves to be completely ineffective in the $\pi^-$D atom, 
so the structure of $V_{\pi^-}$ effectively used here agrees with that of 
$V_{K^-}$, Eqs.~(\ref{eq:V1},\ref{eq:V2}). There is nevertheless one important 
distinction in that the single-nucleon $V_{\pi^-}^{(1)}$ analog of $V_{K^-}^{
(1)}$ is purely real at threshold, so that $\Gamma_{1S}(\pi^-$D) arises 
entirely from the multinucleon $V_{\pi^-}^{(2)}$ analog of $V_{K^-}^{(2)}$. 
Our calculated (calc) $\pi^-$D 1$S$ g.s. level shift and width are 
\begin{equation} 
\epsilon^{\rm calc}_{1S}(\pi^-{\rm D}) = -2.56~{\rm eV}, \,\,\,\,\, 
\Gamma^{\rm calc}_{1S}(\pi^-{\rm D} \to nn) = 0.62~{\rm eV},
\label{eq:pi-th} 
\end{equation} 
disregarding the $\pi^-$D$\,\to nn\pi^0$ partial width which is known to be 
suppressed by several orders of magnitude. Experimentally~\cite{PSI99,PSI10}, 
\begin{equation} 
\epsilon^{\rm exp}_{1S}(\pi^-{\rm D}) = -2.46\pm 0.05~{\rm eV}, \,\,\,\,\, 
\Gamma^{\rm exp}_{1S}(\pi^-{\rm D} \to nn) = 0.86^{+0.03}_{-0.05}~{\rm eV}, 
\label{eq:pi-ex} 
\end{equation}    
where the observed ratio $\Gamma_{\pi^-{\rm D}}(nn)/\Gamma_{\pi^-{\rm D}}
(nn\gamma + nne^+e^-)= 2.76\pm 0.04$~\cite{PSI10} was used. Remarkably, the 
measured $\pi^-$D 1$S$ g.s. level shift is reproduced in this simple model to 
within 4\% (or within 2$\sigma_{\rm exp}$). However, the calculated hadronic 
width amounts to only 72\% of the experimentally determined hadronic width. 
Recalling that $\Gamma_{1S}(\pi^-{\rm D}\to nn)$ is the analog of $\Gamma_{1S}
(K^-{\rm D}\to YN)$ which according to Sect.~\ref{sec:KbarD} contributes 
a few percents at most to the total width $\Gamma_{1S}(K^-$D), one should not 
consider $\Gamma_{1S}(\pi^-{\rm D}\to nn)$ as a test of the optical potential 
approach. Rather, it is only the $\pi^-$D 1$S$ g.s. level shift 
$\epsilon_{1S}(\pi^-$D) that provides in this case a test of 
applying this approach to the $K^-$D 1$S$ g.s. observables.

\section{Remarks on $K^-d$ multiple scattering expansions} 
\label{sec:MS} 

The energy dependence of the $K^-N$ scattering amplitudes below threshold 
was shown in Sect.~\ref{sec:KbarD}, by using two chiral EFT model inputs, 
to increase the width $\Gamma_{1S}(K^-$D) by about 10\%, from 1.1$-$1.2~keV 
to 1.2$-$1.3~keV. A much larger increase, by about 100\% from 1.1~keV to 
2.3~keV, was claimed recently by Liu et al.~\cite{LWLT20} using a similar 
EFT $\bar KN$ amplitudes input. These authors applied an improved version of 
the Deser formula~\cite{BER09} to derive the $K^-$D 1$S$ g.s. level shift and 
width from the $K^-d$ scattering length $a_{K^-d}$. The disagreement noted 
above is traced back to a similar disagreement between respective values of 
Im$\,a_{K^-d}$ considered in these two approaches. From Eq.~(\ref{eq:Ima}) 
here one deduces about 20\% increase of Im$\,a_{K^-d}$, from $\approx$1.2~fm 
to $\approx$1.4~fm, whereas Liu et al.~\cite{LWLT20} find about 70\% increase, 
from $\approx$1.6~fm to $\approx$2.7~fm. 

To figure out the source of such a large increase of Im$\,a_{K^-d}$ in 
Ref.~\cite{LWLT20} we note that these authors rely on extension of the 
fixed-center multiple scattering (MS) expansion of $a_{K^-d}$ in terms of 
$\bar KN$ scattering lengths~\cite{KOR01}, 
\begin{equation}
a_{K^-d}=\frac{\mu_{K^-d}}{m_{K^-}}\int d^3 \vec r \,|\psi_d(\vec r)|^2\, 
{\hat A}_{K^-d}(r) \, , 
\label{eq:akD_static} 
\end{equation} 
where $\psi_d(\vec r)$ is the wave function for nucleons in the deuteron, and 
\begin{equation} 
{\hat A}_{K^-d}(r)=\frac{\tilde a_{K^-p}+\tilde a_{K^-n}+(2\tilde a_{K^-p}
\tilde a_{K^-n}-b_x^2)/r-2b_x^2\tilde a_{K^-n}/r^2}{1-\tilde a_{K^-p}
\tilde a_{K^-n}/r^2+b_x^2\tilde a_{K^-n}/r^3} \, , 
\label{eq:aKD} 
\end{equation} 
with $\tilde a_{\bar K N}=a_{\bar K N}(1+m_K/m_N)$, and 
$b_x= \tilde a_{K^- p\to{\bar K}^0 n}/\sqrt{1+\tilde a_{\bar K^0 n}/r}$. 
This expression can be somewhat simplified if desired~\cite{Gal07}. 

In the first stage Liu et al.~\cite{LWLT20} considered the single-scattering 
(SS) term 
\begin{equation} 
a_{K^-d}^{\rm SS}=\frac{\mu_{K^-d}}{\mu_{K^-N}}\left(\,a_{K^-p}+a_{K^-n}\,
\right), 
\label{eq:SS} 
\end{equation} 
replacing the threshold scattering lengths $a_{K^-N}^{\rm th}$ 
by appropriately constructed subthreshold scattering amplitudes 
$a_{K^-N}^{\rm sub}$ that account for the spectator-nucleon recoil 
in a standard Faddeev approach. This increases the threshold SS value 
Im$\,a_{K^-d}^{\rm SS}$(th)=1.59~fm to Im$\,a_{K^-d}^{\rm SS}$(sub)=2.55~fm 
as listed in the line denoted MS in Table~\ref{tab:MS}. A similar increase 
is observed also in the two other calculations listed in the table, a Faddeev 
calculation by Toker et al.~\cite{TGE81} and the present $K^-$ nuclear optical 
potential calculation. Given that the input threshold values of $a_{K^-p}$ and 
$a_{K^-n}$ in all three calculations are similar, the near agreement between 
their $a_{K^-d}^{\rm SS}$(sub) values is gratifying. 

\begin{table}[htb] 
\caption{$K^-d$ scattering lengths $a_{K^-d}^{\rm full}$ (in fm) 
calculated in three methods: (i) summing up a fixed-scatterer 
multiple-scattering (MS) series~(\ref{eq:aKD}) with $a_{K^-N}^{\rm sub}$ 
input values~\cite{LWLT20}; (ii) solving exact $\bar KNN$ Faddeev equations 
without introducing additional subthreshold dependence~\cite{TGE81}; and 
(iii) solving the $K^-d$ two-body problem using $V_{K^-}$ from a global fit 
to kaonic atoms data, taken here at $\delta\sqrt{s}=-7$~MeV. Single-scattering 
contributions $a_{K^-d}^{\rm SS}$ (\ref{eq:SS}), using $a_{K^-N}^{\rm th}$ and 
$a_{K^-N}^{\rm sub}$ input amplitudes, are also listed.} 
\begin{center} 
\begin{tabular}{ccccc} 
\hline 
Method & Ref. & $a_{K^-d}^{\rm SS}$(th) & $a_{K^-d}^{\rm SS}$(sub) & 
$a_{K^-d}^{\rm full}$ 
\\ 
\hline
MS & \cite{LWLT20} & $-$0.58+i\,1.59 & $-$0.06+i\,2.55 & $-$0.59+i\,2.70 \\ 
Faddeev & \cite{TGE81} & $-$0.37+i\,1.65 & $-$0.16+i\,2.44 & $-$1.47+i\,1.08 \\
$V_{K^-}$ & present & $-$0.08+i\,1.86 & $+$0.18+i\,2.49 & $-$1.26+i\,1.41 \\ 
\hline 
\end{tabular} 
\end{center} 
\label{tab:MS} 
\end{table} 

In the second stage, Liu et al.~\cite{LWLT20} substituted in the MS 
series (\ref{eq:aKD}) the subthreshold amplitudes $a_{K^-N}^{\rm sub}$ 
used in the first stage to construct $a_{K^-d}^{\rm SS}$(sub). 
This increases slightly Im$\,a_{K^-d}^{\rm SS}$(sub)=2.55~fm to 
Im$\,a_{K^-d}^{\rm full}$=2.70~fm, whereas in the full Faddeev 
calculation~\cite{TGE81} Im$\,a_{K^-d}^{\rm SS}$(sub)=2.44~fm decreases 
substantially to Im$\,a_{K^-d}^{\rm full}$=1.08~fm. A similar, 
although somewhat weaker decrease to Im$\,a_{K^-d}^{\rm full}$=1.41~fm 
is observed in our $K^-$ nuclear optical potential $V_{K^-}$ calculation, 
see Eq.~(\ref{eq:Ima}).

\section{Conclusion} 
\label{sec:concl} 

The present note was motivated by a recent claim that the width of the $K^-$D 
atomic 1$S$ g.s. might be larger than 2 keV~\cite{LWLT20}, twice as much as 
values of $\Gamma_{1S}(K^-$D)$\approx\,$1~keV obtained in genuinely three-body 
$K^-$D atom calculations listed in Table~\ref{tab:calcs}. The larger 
$\Gamma_{1S}(K^-$D) is, the more ambiguous the identification of the 1$S$ g.s. 
in forthcoming $K^-$D atom cascade measurements might be~\cite{Cur19}. A width 
$\Gamma_{1S}(K^-$D)$\approx\,$1~keV is expected to ensure the success of these 
experiments. The present calculation of the strong-interaction 1$S$ level 
shift and width in kaonic deuterium uses a $K^-$ nuclear optical potential, 
constructed from chiral-model subthreshold $K^-N$ scattering amplitudes 
$f_{K^-N}$, with added multinucleon dispersion and absorption parameters 
fitted to kaonic atoms data across the periodic table, from $^{7}$Li on. 
This same optical potential methodology has proved successful in pionic 
atoms~\cite{FG14} and was shown in Sect.~\ref{sec:piD} here to reproduce 
remarkably well also the $\pi^-$D atom 1$S$ g.s. level shift, and reasonably 
well its width. 

A key element in securing a good fit to the $K^-$ atoms data is a 
self-consistent implementation of a density dependent $K^-N$ subthreshold 
energy argument $\delta\sqrt{s}$ for $f_{K^-N}$. Here we applied this optical 
potential methodology to the $K^-$D atom using the right-hand side of 
Eq.~(\ref{eq:deltas}) for $\delta\sqrt{s}$, as appropriate to few-body $K^-$ 
atomic and nuclear systems. This gives about 7~MeV downward subthreshold 
shift, considerably less than practised in heavier and denser $K^-$ 
atoms~\cite{FGCHM18}. We then calculated the 1$S$ g.s. level shift and width, 
finding a moderate increase of the width from $\Gamma_{1S}(\delta\sqrt{s}=0)
\approx 1.1-1.2$~keV to $\Gamma_{1S}(\delta\sqrt{s}=-7$~MeV)$\,\approx 
1.2-1.3$~keV. We note that this range of values for $\Gamma_{1S}$ exceeds 
by merely 10-20\% the range of values reached in genuinely three-body 
calculations~\cite{Shev14,Rev16,Hosh17}, see Table~\ref{tab:calcs}. However, 
it is much smaller than the extremely large value $\sim 2.3$~keV reached by 
Liu et al.~\cite{LWLT20} by applying the Deser formula~\cite{BER09} to an 
equally large value of Im$\,a_{K^-d}\sim 2.7$~fm. 

Finally, we compared in Table~\ref{tab:MS} the MS calculation~\cite{LWLT20} of 
Im$\,a_{K^-d}$ to our optical potential calculation, and more significantly to 
a genuinely three-body Faddeev calculation \cite{TGE81}. The extremely large 
value of Im$\,a_{K^-d}^{\rm full}$ in the MS approach cannot be reconciled with 
the moderate values derived in the other two, more conservative approaches. 
We therefore question the validity of using subthreshold amplitudes in the 
fixed-center MS series (\ref{eq:aKD}) to simulate spectator-nucleon recoil 
effects, as done by Liu et al.~\cite{LWLT20}. Recoil corrections were 
considered by Baru et al.~\cite{BER09,MBER15,BEMR15} who concluded that the 
leading recoil effect at threshold contributes less than 10\% to $a_{K^-d}$. 
Forthcoming $K^-$D experiments will hopefully help resolve this issue.

\section*{Appendix A.~~Choice of deuteron density}
\label{sec:appendix} 
\renewcommand{\theequation}{A.\arabic{equation}} 
\setcounter{equation}{0} 
\renewcommand{\thefigure}{A.\arabic{figure}} 
\setcounter{figure}{0} 

Optical model global analyses of hadronic atoms normally exclude atoms 
lighter than Li. Phenomenological nuclear densities are employed in terms of 
experimentally deduced r.m.s. radii. A natural choice for very light atoms is 
harmonic oscillator single-particle (s.p.) nuclear densities which for $A\leq 
4$, namely $s$-shell nuclei, are simply one-parameter Gaussian functions:  
\begin{equation} 
\rho({\vec r}_j;a)=(\sqrt{\pi}a)^{-3}\,\exp\,(-r_j^2/a^2), \,\,\,\,\,\, 
\langle r_j^2 \rangle = \frac{3}{2}\,a^2, \,\,\,\,\,\, j=1,\ldots,A. 
\label{eq:gaussian} 
\end{equation}  
The s.p. density product $\rho({\vec r}_1;a)\ldots\rho({\vec r}_A;a)$ may be 
rewritten in terms of an overcomplete product of a center-of-mass (cm) 
Gaussian $\rho(\vec{\cal R};a/{\sqrt A})$, with $\vec{\cal R}=(1/A)
\sum_{j=1}^{A}{\vec r}_j$, and the same $A$ s.p. Gaussians (\ref{eq:gaussian}) 
with arguments ${\vec r}_j$ replaced by ${\vec r}_j-\vec{\cal R}$. This proves 
useful for $A\gg 4$ when the r.m.s. radius of ${\vec r}_j$ with respect to 
a fixed $\vec{\cal R}$ is directly related to the nuclear matter radius. 

For $A\leq 4$, it is useful to replace the overcomplete relative-coordinate 
density product $\rho({\vec r}_1-\vec{\cal R};a)\ldots\rho({\vec r}_A
-\vec{\cal R};a)$ by a product of $A-1$ Gaussians in terms of Jacobi relative 
coordinates and their related size parameters $a_i$:  
\begin{equation} 
{\vec{\cal R}}_i = {\vec r}_i - \frac{1}{i-1}\sum_{j=1}^{i-1}{\vec r}_j, 
\,\,\,\,\,\, a_i=\sqrt{\frac{i-1}{i}}a,\,\,\,\,\,\,i=2\ldots A.  
\label{eq:Jacobi} 
\end{equation} 
Here, the last Jacobi coordinates ${\vec{\cal R}}_A$ is proportional 
to the s.p. coordinate ${\vec r}_A$ with respect to the cm coordinate 
${\vec{\cal R}}$: 
\begin{equation} 
{\vec {\cal R}}_A = \frac{A}{A-1}({\vec r}_A - \vec{\cal R}).  
\label{eq:last} 
\end{equation} 
To relate the desired $\langle {\cal R}_A^2 \rangle$ to $\langle 
({\vec r}_A - \vec{\cal R})^2 \rangle$ we recall the $\frac{A}{A-1}$ factor 
in (\ref{eq:last}) and the $\sqrt{\frac{A-1}{A}}$ factor for $i=A$ in 
(\ref{eq:Jacobi}), yielding 
\begin{equation} 
\langle {\cal R}_A^2 \rangle = \frac{A}{A-1}\,\langle 
({\vec r}_A - \vec{\cal R})^2 \rangle,  
\label{eq:last+} 
\end{equation} 
in agreement with Elton's book~\cite{Elton}. For the deuteron, $A$=2, 
subtracting the proton charge radius squared from the deuteron charge radius 
squared as given in Ref.~\cite{ADNDT13} we obtain a matter r.m.s. radius of 
1.954~fm, which upon multiplying by ${\sqrt 2}$ from (\ref{eq:last+}) gives 
${\langle {\cal R}_{A=2}^2 \rangle}^{1/2}$=2.763~fm, as used in the present 
meson-deuteron calculations.

\section*{Acknowledgements} 

We thank Zhan-Wei Liu and Wolfram Weise for useful remarks on the previous 
version. The work of NB was supported by the PAZY Foundation and by the ISF 
grant No. 1308/16. The work of all three authors is supported by the European 
Union's Horizon 2020 research and innovation programme under grant agreement 
No. 824093.

\end{document}